\documentclass{article}

\usepackage{arxiv}

\usepackage[utf8]{inputenc} 
\usepackage[T1]{fontenc}    
\usepackage{hyperref}       
\usepackage{url}            
\usepackage{booktabs}       
\usepackage{amsfonts}       
\usepackage{nicefrac}       
\usepackage{microtype}      
\usepackage{float}     

\usepackage{subcaption} 

\usepackage{lipsum}         
\usepackage{amsmath}        
\usepackage{graphicx}
\usepackage{subcaption}

\usepackage{natbib}
\usepackage{cleveref}       
\usepackage{doi}
\usepackage[font=small,labelfont=bf]{caption}
\usepackage{xcolor}

\title{Preliminary analysis of Sus scrofa movement using hidden Markov models and Networks}

\definecolor{RossoE}{RGB}{198, 0, 0}%

\newif\ifuniqueAffiliation
\uniqueAffiliationfalse  
\ifuniqueAffiliation 
\author{
}
\else
\usepackage{authblk}

\newbox{\orcid}\sbox{\orcid}{\includegraphics[scale=0.06]{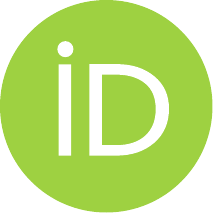}} 

\author[1,3,4]{%
    \href{https://orcid.org/0009-0002-3157-1749}{\usebox{\orcid}\hspace{1mm}Riccardo Basilone}%
}

\author[2]{%
    \hspace{1mm}Eleonora Bergamin%
}

\author[1,3,4]{%
    \href{https://orcid.org/0009-0001-2517-941X}{\usebox{\orcid}\hspace{1mm}Federica Fanelli}%
}

\author[5,6]{%
    \href{https://orcid.org/0000-0001-6690-5345}{\usebox{\orcid}\hspace{1mm}Egor Kotov}%
}

\author[7,8]{%
\href{https://orcid.org/0000-0003-3143-8075}
{\usebox{\orcid}\hspace{1mm}Kevin Morelle}%
}

\author[9]{%
    \href{https://orcid.org/0009-0003-6501-2520}{\usebox{\orcid}\hspace{1mm}Alisa Klamm}%
}

\author[10,11,12]{%
    \href{https://orcid.org/0009-0004-0275-6677}{\usebox{\orcid}\hspace{1mm}Manal Nhili}%
}

\author[13]{%
    \href{https://orcid.org/0009-0001-3646-8399}{\usebox{\orcid}\hspace{1mm}Joshua Rosen}%
}

\author[14]{%
    \href{https://orcid.org/0000-0003-4216-5746}{\usebox{\orcid}\hspace{1mm}Andrew Schendl}%
}

\author[15]{%
    \href{https://orcid.org/0000-0002-6138-7902}{\usebox{\orcid}\hspace{1mm}Olena  Holubowska}%
}

\author[16]{%
    \href{https://orcid.org/0000-0002-9902-9035}{\usebox{\orcid}\hspace{1mm}Andrew Renninger}%
}

\author[17]{%
    \href{https://orcid.org/0000-0001-9113-6090}{\usebox{\orcid}\hspace{1mm}Kamil Smolak\thanks{Correspondence: \texttt{kamil.smolak@upwr.edu.pl}}}%
}

\affil[1]{Department of Physics, Sapienza University, Rome, Italy}
\affil[2]{Scuola Internazionale Superiore di Studi Avanzati, Trieste, Italy}
\affil[3]{Sony Computer Science Laboratories - Rome, Joint Initiative CREF-Sony, Centro Ricerche Enrico Fermi, Rome, Italy}
\affil[4]{Centro Ricerche Enrico Fermi, Rome, Italy}
\affil[5]{Max Planck Institute for Demographic Research, Rostock, Germany}
\affil[6]{Universitat Pompeu Fabra, Barcelona, Spain}
\affil[7]{Max Planck Institute of Animal Behavior, Konstanz, Germany}
\affil[8]{Minuartia, Barcelona, Spain}
\affil[9]{Upper Danube Nature Conservation Center, Bad Langensalza, Germany}
\affil[10]{ENSIAS, Mohammed V University, Rabat, Morocco}
\affil[11]{UMR ASTRE, CIRAD, Montpellier, France}
\affil[12]{UMR ASTRE, University of Montpellier, Montpellier, France}
\affil[13]{School of Public Policy and Urban Affairs, Northeastern University, Boston, MA, USA}
\affil[14]{School of Environmental and Natural Sciences, Bangor University, Bangor, UK}
\affil[15]{Department of Earth and Environmental Sciences, KU Leuven, Leuven, Belgium}
\affil[16]{Centre for Advanced Spatial Analysis, University College London, London, UK}
\affil[17]{Faculty of Environmental Engineering and Geodesy, Wrocław University of Environmental and Life Sciences, Wrocław, Poland}

\fi

\hypersetup{
pdftitle={Preliminary analysis of Sus scrofa movement using Hidden Markov Models and Networks},
pdfsubject={q-bio.PE, q-bio.QM},
pdfauthor={Ricardo Basilone, Eleonora Bergamin, Federica Fanelli, Egor Kotov, Kevin Morelle, Alisa Klamm, Manal Nhili, Joshua R. Rosen, Andrew Schendl, Olena  Holubowska, Andrew Renninger, Kamil Smolak},
pdfkeywords={complex networks, animal movements, attractors, repellors},
}

\begin{document}

\maketitle

\begin{abstract}
This study examines the complex movement patterns and behavioral characteristics of wild boars using GPS telemetry data collected over a two-month period. Our methodological approach centers on the application of an autoregressive hidden Markov model (AR-HMM) to discern distinct behavioral states embedded within the trajectories. Furthermore, the study aimed to construct behavioral networks, derived from these segmented trajectories. The resultant network structures showed that the hidden behavioral patterns are mostly independent of geographical locations. While most locations have many behaviors occurring in them, our findings also suggest that individual boar behaviors vary in spatial expansiveness. Finally, the research incorporates a spatial trajectory analysis, complemented by raster data validation, to potentially delineate areas acting as repellents within the ecological context of Hainich National Park in Germany. 
\end{abstract}

\keywords{Complex Networks \and Animal Movements \and Hidden Markov Model \and Eurasian Wild Boar}

\section{Introduction}
Understanding animal movement patterns is crucial for ecological research, conservation planning, and wildlife management, particularly for species that significantly impact ecosystems and human activities. This study examines the movement ecology of Eurasian wild boars (\textit{Sus scrofa}), a species whose expanding populations, broad habitat use, and economic impact have put it in the cross-hairs of wildlife management across Europe \citep{Massei2015Apr}. Movement is a fundamental ecological process that connects individuals with their environment, with broader impacts on population dynamics, species distribution, and ecosystem functioning \citep{Nathan2008Dec}. Movement ecology is an integrative framework that seeks to understand these mechanisms of movement, incorporating four key components: the internal state of the individual, its motion capacity, navigational capacity, and external factors of the environment \citep{Nathan2008Dec}. This framework has been applied across diverse taxa, from insects to large mammals, with much of the literature examining how environmental variables, the landscape, and anthropogenic factors shape their movement patterns\citep{Joo2022Dec}. Modern movement ecology research has applied advanced fine-scale tracking methods using GPS tags, accelerometers, and remote sensing, which have provided greater insight into the movement behaviors and spatial patterns of various species \citep{Williams2019Oct}. Recent methodological advancements include the application of hidden Markov models (HMMs) to uncover behavior-dependent resource selection patterns to understand movement causation \citep{McClintock2020Oct, Michelot2016May}. 

Wild boars represent a vital species for movement ecology research due to their adaptability to urban and rural environments, the risk of disease transmission, and the significant ecological and economic impacts their foraging behavior can cause. Since the 1980s, wild boar populations have proliferated throughout Europe, with some regions, such as Sweden, experiencing fluctuations from near-extinction to super-abundance within just 35 years \citep{Massei2015Apr, Bergqvist2024Dec}. This population explosion, combined with their ability to cause substantial agricultural damage and risk of spreading disease such as African Swine Fever (ASF) to domestic livestock \citep{Sauter-Louis2021Aug, Massei2015Apr}, has made understanding their movement patterns and habitat preferences a critical priority for wildlife management and conservation planning. 

Wild boars exhibit complex spatial behavior characterized by their strong olfactory sense and advanced cognitive abilities, which enable sophisticated responses to environmental cues \citep{Morelle2015Jan}. Home range sizes vary dramatically across populations and habitats, with mean values ranging between 4 and 11 km$^2$ (95 percent kernel density estimation) or 21-52 km$^2$ (minimum convex polygon estimation) depending on the modeling approach and environmental context \citep{Miettinen2023Oct}. The species demonstrates strong site fidelity upon reaching maturity, with most individuals dispersing only 1-2 km from their birth range throughout their lives, while only a smaller number become long-distance travelers, extending their range to 20-30 km \citep{podgorskiWildBoarMovements2018}. Adult males typically maintain larger home ranges and have higher activity levels than females, with spatial parameters showing significant variation between hunting-vulnerable areas and protected environments \citep{Laguna2021Nov}. Daily movement patterns are predominantly nocturnal, with activity patterns varying significantly between human-disturbed and natural environments. 

Daily movements of wild boars are primarily driven by resource availability and quality, with individuals capable of traveling several kilometers to access preferred food sources and environmental conditions. Using HMMs, three distinct behavior patterns have been previously identified in wild boar during their daily movement: resting, foraging, and traveling \citep{Clontz2021Mar}. Water sources represent one of the strongest attractors, particularly during hot summer months \citep{Singer1981}. Boars demonstrate a strong preference for areas within 500 meters of permanent water bodies, with movement paths often following corridors that provide both water and protection \citep{Stillfried2017Dec}. Agricultural fields serve primarily as repellents, being actively avoided throughout most of the year due to the open exposure presented by the landscape. However, during the summer months, these areas serve as attractors, with boars increasing the distances they travel in order to exploit the rich food source \citep{Laguna2021Nov}. There has been an attempt to restrict wild boar movement through the introduction of physical repellents such as odor fences \citep{Faltusova2024Sep} and ungulate-proof fencing \citep{Koriakin2024Jul}, with the latter only proving to be marginally effective at controlling the movement of these highly adaptable species. 

As we noted previously, HMMs are among the state-of-the-art tools in movement ecology \citep{Glennie2022Feb}, enabling the identification of distinct behavioral states from movement data to better understand the drivers behind animal behavior. The adoption of HMMs in movement ecology is in large part thanks to the wide availability of software such as \texttt{R} packages \texttt{moveHMM} \citep{michelot.langrock.ea2016}, \texttt{momentuHMM} \citep{mcclintock.michelot2018}, and \texttt{Python} libraries such as \texttt{smm} \citep{Linderman_SSM_Bayesian_Learning_2020}. While HMMs have proven effective at segmenting movement trajectories into discrete behavioral states \citep{Conners2021Dec}, they exhibit several limitations that restrict their applicability to complex ecological systems. Standard HMMs assume that behavioral transitions depend only on the current state, without considering the animal's movement history \citep{Glennie2022Feb}. 

The autoregressive hidden Markov model (AR-HMM) \citep{box_time_1994, shumway_time_2000} is particularly well-suited for classifying or segmenting wildlife trajectories like those of wild boars because it extends standard HMMs by incorporating movement history into the observation process.
Although AR-HMMs are not able to model a continuous latent state (as in the case of Switching Linear Dynamical Systems instead \cite{box_time_1994, shumway_time_2000}), they are able to implicitly capture behaviors that are state-dependent and autoregressive, i.e., depending on the past observations. This allows AR-HMMs to describe context-dependent switching without requiring the explicit modeling of the environment changes that can affect the animals' behavior.

AR-HMMs can be considered a specific instance of SLDS in which the latent state is a deterministic function of the past observations \citep{fox_nonparametric_2008, linderman.johnson.ea}. As such, they retain the interpretability and scalability of HMMs while better capturing temporal dependencies in movement data. Although less expressive than full SLDS models since they lack an explicit continuous latent dynamics, AR-HMMs represent a practical balance for ecological applications where discrete behavioral states are primary, but movement history matters.

Network approaches offer a powerful framework for understanding animal-place interactions where locations serve as nodes and movements or interactions form edges. Within movement ecology, there are several methodological approaches for creating location networks from GPS tracking data, with behavior segmentation providing crucial data for meaningful network construction \citep{Bastille-Rousseau2018Apr}. Combining HMMs with the networks to first identify behavioral states, then apply clustering methods to define location nodes, has been proven to be a robust approach \citep{Joo2020Jan}. These networks allow for the quantification of habitat usage patterns through standard network metrics, including degree centrality, which identifies frequently visited locations, and betweenness centrality, which reveals movement corridors that identify habitat connections \citep{Sosa2021Jan}. Alternative network approaches include temporal motif analysis, which transforms trajectories into spatial networks while preserving the temporal movement sequences as three-node subgraphs \citep{Pasquaretta2021Jan}. Social network approaches have also been adapted for movement analysis, creating proximity-based networks where edges represent spatial or temporal co-occurence of individuals, enabling the study of collective movement patterns \citep{Farine2015Sep}.

This paper takes an established hidden Markov model (HMM) framework for animal movement and adapts it to infer latent behavioral states—like foraging, resting, and mating, based on directed movement from mobility traces of wild boars. It then uses those inferred states to build a weighted bipartite network, linking each boar to the behaviors it most frequently occupies and revealing clusters of individuals with similar movement repertoires. We overlay these state‐specific location densities onto high‐resolution raster layers (e.g., land cover, elevation, precipitation) to statistically identify attractor zones and repellent areas. Our findings highlight movement corridors and habitat hotspots where ASF transmission may occur, quantifying the structure of boar behavioral networks in space. We place this work within a broader effort to understand, simulate, and ultimately predict boar movement dynamics-critical for targeting interventions to curb disease spread and manage invasive populations.

\section{Methods}
\label{sec:Methods}

\subsection{Data preprocessing and analysis}
\label{sec:data}
We used high-frequency Global Positioning System (GPS) data representing the movement of Eurasian wild boar. Data were collected in the vicinity of Hainich National Park, located in western Thuringia, central Germany, between October $1^{\text{st}}, \ 2016$, and December $31^{\text{st}}, \ 2019$. Hainich National Park includes a "no-hunting zone" that covers $33 \ \text{km}^2$. The total park area is $75 \ \text{km}^2$ \citep{wielgus2023first}. 

Data were collected from 57 ASF-free individuals, captured at four different locations within the national park, close to the no-hunt zones. Each boar was tagged with a Vectronic Aerospace GPS-GSM Vertex Lite collar, which determines its position via GPS and transmits the data through the GSM network to a central database. All individuals were released immediately after collaring. The captured wild boars varied in weight (30 to 80 kg) and age, including juveniles, yearlings, and adults.

Each individual has a raw movement trajectory \( T_i \), defined as:
\[
T_i = \{(l_1, t_1), (l_2, t_2), \dots, (l_n, t_n)\}
\]
where \( l_j = (x_j, y_j) \) are the geographical coordinates of the \( j \)-th data point, \( t_j \) is the timestamp associated with that data point, and \( n \) is the total number of recorded positions.

We preprocessed the dataset to ensure high quality and maximize data coverage, following commonly applied data filtration methodologies from previous studies \citep{song2010limits, smolak2021impact}. Specifically, we applied selection criteria based on data completeness and trajectory duration. We define completeness \( q \) as the fraction of unpopulated hourly intervals, that is, intervals where no data point is present. In human mobility research, it has been demonstrated that approximately 14 days of observation are typically sufficient to capture the routine mobility patterns of most individuals. Since no such threshold has yet been established for animal mobility, we adopted a conservative approach and retained only individuals whose trajectories were continuously recorded for at least 20 days, with a completeness \( q < 0.6 \). After filtering, the final data set consisted of 31 individuals, each observed for an average of 21 days, with an average completeness of \( q = 0.09 \pm 0.2 \). This level of data coverage is considered sufficient to avoid biases in mobility statistics \citep{lin2012predictability}. The observations span the period from October to December in both 2017 and 2018, which coincides with the breeding season for wild boars.

\begin{figure}[!ht]
  \centering
  \includegraphics[scale=0.6]{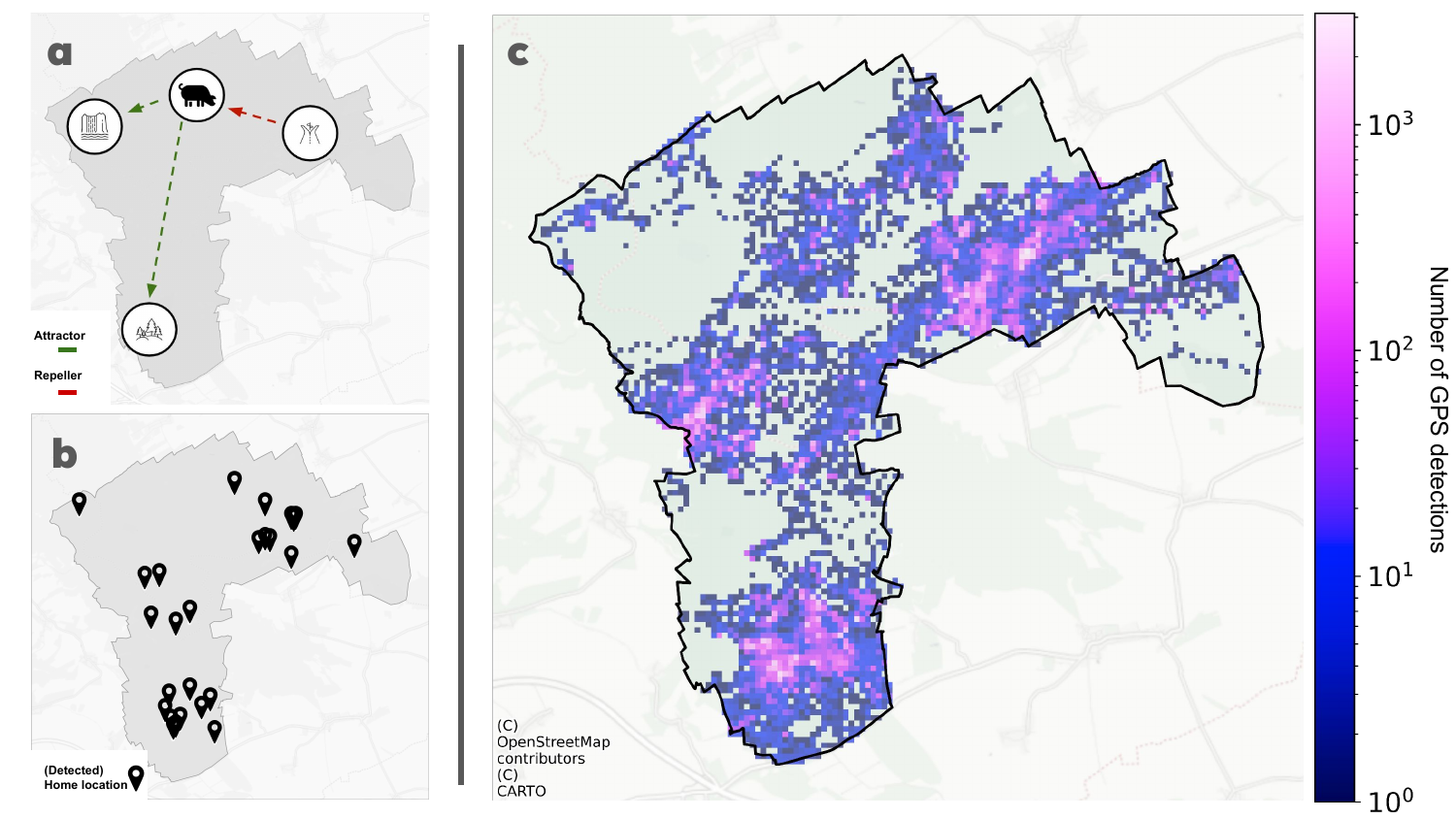}
  \caption{Wild boar movement summary.  
    (\textbf{a}) Stylized diagram of boar mobility flows between detected \emph{stay locations}.  
    (\textbf{b}) Geographic map of detected boar home locations within Hainich National Park.  
    (\textbf{c}) Heat map of boar flow frequency between \emph{stay locations} within Hainich National Park, with warmer colors indicating higher traffic. Note: Stay locations outside of the national park boundary were excluded for visualization purposes.} 
  \label{fig:boar-movement}
\end{figure}

We further processed the filtered dataset to detect \textit{stay locations}, where individuals stop to engage in activities. This process effectively segments the trajectories into ``moving'' and ``resting'' states. Figure \ref{fig:boar-movement} highlights the density of flows between the stay locations. To detect stay locations, we applied the widely used two-step algorithm from \citep{hariharan2004project}. In the first step, the algorithm iterates through each trajectory in chronological order, identifying all locations within a roaming range \( \delta \) of each other as potential stay locations. For each candidate, the algorithm checks whether the individual remained within that location for at least \( \tau \) minutes. If so, the candidate is confirmed as a stay location; otherwise, it is discarded. The algorithm then continues iterating through the remaining trajectory. In the second step, the identified stay locations are clustered to group nearby locations under the same label, effectively recognizing that these stay locations represent the same physical place. For this clustering step, we used the Density-Based Spatial Clustering of Applications with Noise (DBSCAN) algorithm, controlled by the \( \epsilon \) parameter \cite{ester1996density}.

To select appropriate algorithm parameters, we performed a sensitivity analysis to identify parameter ranges where the number of detected stay locations remained stable and invariant to small parameter changes. This stability indicates that the detected stay locations are well separated and robust to the specific parameter choice. Based on this analysis, we selected the roaming range: \( \delta = 80 \) meters, minimum duration \( \tau = 10 \) minutes, and DBSCAN clustering radius: \( \epsilon = 170 \) meters.

We performed an initial data analysis using mobility metrics widely adopted in human mobility science \citep{barbosa2018human}. We applied individual-level metrics that capture both spatial and temporal movement characteristics. These include: the distribution of waiting times, computed as the time an individual remains in a stay location; jump length distributions, where the jump length is defined as the distance covered by individuals between consecutively visited locations; distinct locations over time, which examine the growth in the number of unique locations visited over time, reflecting exploration dynamics; and visitation frequency, which measures the normalized frequency of visits to each identified location, providing insight into space use heterogeneity. We also examined the entropy-based framework proposed by \citep{song2010limits}, which provides a theoretical upper bound on how predictable an individual’s movement is, based on the diversity and frequency of visited locations.

\subsection{Autoregressive hidden Markov model}
The Autoregressive hidden Markov model implementation follows what is described in \citep{fox_nonparametric_2008} and \citep{linderman.johnson.ea}.

\begin{figure}[!ht]
  \centering
  \includegraphics[width=0.8\linewidth]{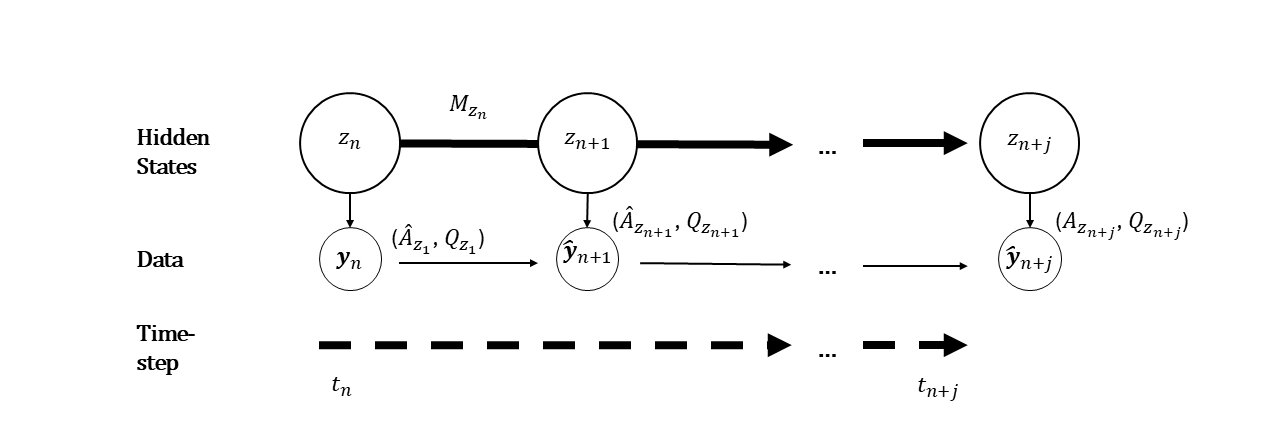}
  \caption{ \small Plate diagram of the autoregressive hidden Markov model (AR-HMM). Larger circles denote the hidden discrete mode variables $z_t$, which evolve according to a $K\times K$ transition matrix $M$ with entries $M_{ij}=P(z_{t+1}=j\mid z_t=i)$. Smaller circles denote the observed data $y_t$, which in the collapsed AR(1) formulation satisfy $y_t = A_{z_t}y_{t-1} + b_{z_t} + v_t$, where $v_t\sim\mathcal{N}(0,Q_{z_t})$.  Each arrow $y_{t-1}\to y_t$ is labeled by the mode-dependent pair $(A_{z_t},Q_{z_t})$.}
  \label{fig:arslds-diagram}
\end{figure}

The switching linear system is described as follows. At each time step $t \in \{ 1, ..., T\}$ the system is in a latent state $z_t \in \{ 1, ..., K\}$ and in a certain configuration given by $\textbf{y}_t \in \mathbb{R}^N$. 
Latent state is updated every time through a Markovian process. The generative model is indeed ruled by the following equations:

\begin{equation}
    z_{t+1} \sim \mathrm{Cat}(\mathbf{M}_{z_{t}}), \quad \quad
    \mathbf{v}_{z_{t+1}} \sim \mathcal{N}(\mathbf{0}, Q_{z_{t+1}}), \quad \quad
    \mathbf{y}_{t+1} = \hat{A}_{z_{t+1}} \hat{\mathbf{y}}_t + \mathbf{v}_{z_{t+1}}
\end{equation}

where $M \in \mathbb{R}^{K \times K}$ is the Markovian transition matrix ( hence $\textbf{M}_{k}$ its $k$-th row), $Q_k \in \mathbb{R}^{N \times N}$ are the covariance matrices for the Gaussian noise associated with each state, $A_k \in \mathbb{R}^{N \times N}$ and $\textbf{b}_k \in \mathbb{R}^{N}$ are the parameters of the linear model for each state. For simplicity, the last can be grouped in a single matrix $\hat{A}_{k} = [\textbf{b}_{k}, A_{k}] \in \mathbb{R}^{N \times (N+1)}$, and the spatial coordinate $\textbf{y}_t$ is augmented as $\hat{\textbf{y}}_t = [1, \textbf{y}_t] \in \mathbb{R}^{N+1}$ (see Figure \ref{fig:arslds-diagram}).

The likelihood of a certain trajectory $\textbf{y}_{1:T}$, given the latent states $z_{1:T}$ and all model parameters, can be written in two different formulations:

\begin{equation}
\begin{aligned}
    P(\textbf{y}_{1:T}|z_{1:T}, \{\hat{A}_k\}, \{Q_k\}, M) & = \prod_{t=1}^{T-1} (2\pi)^{-\frac{N}{2}} |Q_{z_{t+1}}|^{-\frac{1}{2}}
    \cdot \exp{\left[-\frac{1}{2} (\textbf{y}_{t+1} - \hat{A}_{z_{t+1}} \hat{\textbf{y}}_t)^T Q_{z_{t+1}}^{-1} (\textbf{y}_{t+1} - \hat{A}_{z_{t+1}} \hat{\textbf{y}}_t) \right]} \\ & = \prod_{k=1}^{K} (2\pi)^{-\frac{N \cdot N_k}{2}} |Q_{k}|^{-\frac{N_k}{2}}
    \cdot \exp{\left[-\frac{1}{2} \sum_{t_i \in k}^{N_k} (\textbf{y}_{t_i} - \hat{A}_{k} \hat{\textbf{y}}_{t_i-1})^T Q_{k}^{-1} (\textbf{y}_{t_i} - \hat{A}_{k} \hat{\textbf{y}}_{t_i-1}) \right]}
\end{aligned}
\end{equation}

where $t_i$ is an instant in time in which $z_{t_i} = k$ and $N_k$ corresponds to the number of occurrences of state $k$.

The joint probability distribution for all the variables can be written as

\begin{equation}
\begin{aligned}
    P(\textbf{y}_{1:T}, z_{1:T}, \{\hat{A}_k\}, \{Q_k\}, M) =
    P(\textbf{y}_{1:T}|z_{1:T}, \{\hat{A}_k\}, \{Q_k\}, M) \cdot P(z_{1:T}|M) \cdot P(M) \cdot P(\{\hat{A}_k\}, \{Q_k\}) ,
\end{aligned}
\end{equation}

where choosing an uninformative prior on the state $z_1$ the probability of observing a given latent history is

\begin{equation}
    P(z_{1:T}| M) = P(z_1) \cdot \prod_{t=1}^{T-1} M_{z_t, z_{t+1}} = \frac{1}{K} \cdot \prod_{t=1}^{T-1} M_{z_t, z_{t+1}}. 
\end{equation}

Conjugate priors for the model parameters can be chosen, in order to simplify the inference task. Specifically, the Dirichlet distribution is the conjugate prior of the categorical distribution, and the Inverse Wishart ($\mathcal{IW}$) distribution is the conjugate prior for the covariance matrix of a Multivariate Normal ($\mathcal{MN}$) distribution.

\begin{equation}
    P(M) = \prod_{k=1}^{K} P(\textbf{M}_k|\boldsymbol\alpha_k) = \prod_{k=1}^{K} \frac{1}{B(\boldsymbol\alpha_k)} \prod_{l=1}^{K} M_{kl}^{\alpha_{kl} - 1} , 
\end{equation}
\begin{equation}
\begin{aligned}
    P(\{\hat{A}_k\}, \{Q_k\}) & = \prod_{k=1}^{K} P(\hat{A}_k, Q_k) = \prod_{k=1}^{K} \mathcal{MNIW} (\hat{A}_k, Q_k; C, V, S, \nu) \\ & = \prod_{k=1}^{K} \mathcal{MN}(\hat{A}_k| C, Q_k, V) \cdot \mathcal{IW}(Q_k; S, \nu) .
\end{aligned}
\end{equation}

The model hyperparameters choice has to reflect the knowledge about the system. We use hence uninformative priors. The concentration parameters of the Dirichlet distribution are uniformly set as \(\boldsymbol\alpha_k = [1, ..., 1]\ \ \forall k = 1, ..., K\),  while the mean matrix $ C $ takes the form $[\vec{\mathbf{0}}, \mathbb{I}_N]$.  The inverse covariance matrix along columns, $ V $, is assigned the structure $\mathbb{I}_{N+1} $.  and the scale matrix $ S = \mbox{diag}(\lambda_1, ..., \lambda_N)$ associated with the Inverse Wishart ($\mathcal{IW}$) distribution, defines the expected scale of the data variability. The degrees of freedom for the $\mathcal{IW}$ distribution are set to $\nu = N$, matching the data dimensionality. Collectively, the hyperparameters consist of the  matrices $\alpha$, $C$, $V$, $S$ and the scalar number of degrees of freedom $\nu$.

The objective is then to perform Gibbs sampling for this kind of AR-HMM. We have to compute the conditional probabilities of each random variable, keeping fixed all the others.

\subsubsection{Message passing algorithm on $z_{1:T}$}
Following \citep{fox_nonparametric_2008}, one can develop a message passing algorithm to sample the hidden history $z_{1:T}$. First, we observe that the joint probability $P(z_{1:T}| \{\hat{A}_k\}, \{Q_k\}, M, \textbf{y}_{1:T})$ factorizes as

\begin{equation}
    P(z_{1:T}| \{\hat{A}_k\}, \{Q_k\}, M, \textbf{y}_{1:T}) = P(z_1| \{\hat{A}_k\}, \{Q_k\}, M, \textbf{y}_{1:T}) \cdot \prod_{t = 2}^{T} P(z_t|z_{t-1}, \{\hat{A}_k\}, \{Q_k\}, M, \textbf{y}_{1:T}).
\end{equation}

Starting from sampling $z_1$ then one can sample $z_2|z_1$, and so on.
Single terms of such product can be iteratively computed from the message vector $\mathbf{m}_{t, t-1}$, whose $k$-th component is $m_{t, t-1}(z_{t-1} = k)$ defined as

\begin{equation}
    m_{t, t-1}(z_{t-1} = k) \propto
    \begin{cases}
        \sum_{j=1}^{K} M_{k, j} \cdot \mathcal{N}(\textbf{y}_{t} - \hat{A}_{j} \hat{\textbf{y}}_{t-1}, Q_{j}) \cdot m_{t+1, t}(z_{t} = j) \ \ \ t \leq T , \\
        1 \ \ \ \ \ \ \ \ \ \ \ \ \ \ \ \ \ \ \ \ \ \ \ \ \ \ \ \ \ \ \ \ \ \ \ \ \ \ \ \ \ \ \ \ \ \ \ \ \ \ \ \ \ \ \ \ \ \ \ \ \ \ \ \ \ t = T+1 .
    \end{cases}
\end{equation}

The conditional distribution of $z_1$ corresponds to

\begin{equation}
    P(z_{1}| \{\hat{A}_k\}, \{Q_k\}, M, \textbf{y}_{1:T}) \propto P(z_1) P( \textbf{y}_1| \{\hat{A}_k\}, \{Q_k\}) \sum_{z_{2:T}} \prod_{t} P(z_t| M_{z_{t-1}})P( \textbf{y}_t| \{\hat{A}_k\}, \{Q_k\}, \textbf{y}_{t-1}) \, .
\end{equation}

Supposing to sample $z_1$ from a uniform categorical distribution, all the others $z_{t}$ can be sampled categorically from

\begin{equation}
    P(z_t = k|\{\hat{A}_k\}, \{Q_k\}, M, \textbf{y}_{1:T}) \propto M_{z_{t-1}, k} \cdot \mathcal{N}(\textbf{y}_{t} - \hat{A}_{k} \hat{\textbf{y}}_{t-1}, Q_{k}) \cdot m_{t+1, t}(z_{t} = k) .
\end{equation}

\subsubsection{Conditional distributions on $M$, $\{\hat{A}_k\}, \{Q_k\}$}

In order to sample the model parameters $M$, $\{\hat{A}_{k}\}$, $\{Q_{k}\}$, conditional distribution are found to follow the given proportionalities \citep{fox_nonparametric_2008}.

\begin{equation}
\begin{aligned}
    P(\textbf{M}_{k} | z_{1:T}, \{\hat{A}_k\}, \{Q_k\}, \textbf{y}_{1:T}) & \propto \prod_{t=1}^{T-1} \chi(z_t = k) \cdot M_{k, z_{t+1}} \cdot \prod_{l=1}^K M_{kl}^{\alpha_{kl} - 1} \\ & \propto \prod_{l=1}^K M_{kl}^{\alpha_{kl} + \sum_t^{T-1} \chi(z_t = k) \chi(z_{t+1} = l)\ - 1} \\ & \propto \mbox{Dir}(\textbf{M}_{k} | \boldsymbol \alpha_k + \boldsymbol \chi_k) 
\end{aligned}
\label{eq:meq}
\end{equation}

Following \citep{fox_nonparametric_2008} we also write the likelihood in matrix normal form. We build the matrices $Y^{(k)} = [\textbf{y}_{t_1}, ..., \textbf{y}_{t_{N_k}}] \in \mathbb{R}^{N \times N_k}$ and $\bar{Y}^{(k)} = [\hat{\textbf{y}}_{t_1 - 1}, ..., \hat{\textbf{y}}_{t_{N_k} - 1}] \in \mathbb{R}^{(N+1) \times N_k}$ for this goal. We obtain then

\begin{equation}
\begin{aligned}
    P(\textbf{y}_{1:T}|z_{1:T}, \{\hat{A}_k\}, \{Q_k\}, M) & = \prod_{k=1}^{K} (2\pi)^{-\frac{N \cdot N_k}{2}} |Q_{z_{t+1}}|^{-\frac{N_k}{2}} \\ & \cdot \exp{\left[-\frac{1}{2} \mbox{tr}\left( (Y^{(k)} - \hat{A}_{k} \bar{Y}^{(k)})^T Q_{k}^{-1} (Y^{(k)} - \hat{A}_{k} \bar{Y}^{(k)}) \mathbb{I}_{N_k} \right) \right]} \\ & = \prod_{k=1}^{K} \mathcal{MN}(Y^{(k)}; \hat{A}_{k} \bar{Y}^{(k)}, Q_k, \mathbb{I}_{N_k})
\end{aligned}
\end{equation}

Since the prior for $\hat{A}_{k}$ follows a matrix normal distribution, its conditional posterior also remains a matrix normal distribution, but with updated parameters. Defining the following quantities

\begin{equation}
    S_{\bar{y} \bar{y}}^{(k)} = \bar{Y}^{(k)} \bar{Y}^{(k) T} + V, \quad
    S_{y \bar{y}}^{(k)} = Y^{(k)} \bar{Y}^{(k) T} + CV,
\end{equation}
\begin{equation}
    S_{y y}^{(k)} = Y^{(k)} Y^{(k) T} + CVC^T, \quad
    S_{y | \bar{y}}^{(k)} = S_{y y}^{(k)} - S_{y \bar{y}}^{(k)} S_{\bar{y} \bar{y}}^{(k) -1} S_{y \bar{y}}^{(k) T}
\end{equation}

the conditional probabilities for the linear dynamics parameters are

\begin{equation}
    P(Q_k | z_{1:T}, \hat{A}_k, M, \textbf{y}_{1:T}) \propto \mathcal{IW}(Q_k| S + S_{y | \bar{y}}^{(k)}, \nu + N_k)
\label{eq:qeq}
\end{equation}
\begin{equation}
    P(\hat{A}_k | z_{1:T}, Q_k, M, \textbf{y}_{1:T}) \propto \mathcal{MN}(\hat{A}_k | S_{y \bar{y}}^{(k)} S_{\bar{y} \bar{y}}^{(k) -1}, Q_k, S_{\bar{y} \bar{y}}^{(k) -1})
\label{eq:aeq}
\end{equation}

Equations \eqref{eq:meq}, \eqref{eq:qeq}, \eqref{eq:aeq} allow then to sample the model parameters $M$, $\{\hat{A}_{k}\}$, $\{Q_{k}\}$.

\subsubsection{Specifics of the experiments}

Following the literature for animal movement modeling \cite{erdtmann_behavioural_2020, Clontz2021Mar, Laguna2021Nov, saldanhaAnimalBehaviourMove2023}, we set the number of hidden states to $K = 5$. We analyze each boar individually, defining its configuration at time $t$ as a two-dimensional vector $\textbf{y}_t \in \mathbb{R}^2 = \{ x_t, y_t \}$ representing its spatial coordinates. For each boar, the complete trajectory is then given by the sequence of all its subsequent states across all observed timesteps.

\begin{figure}[H]
  \centering
  \begin{minipage}[c]{0.45\textwidth}  
    \centering
    \includegraphics[width=\linewidth]{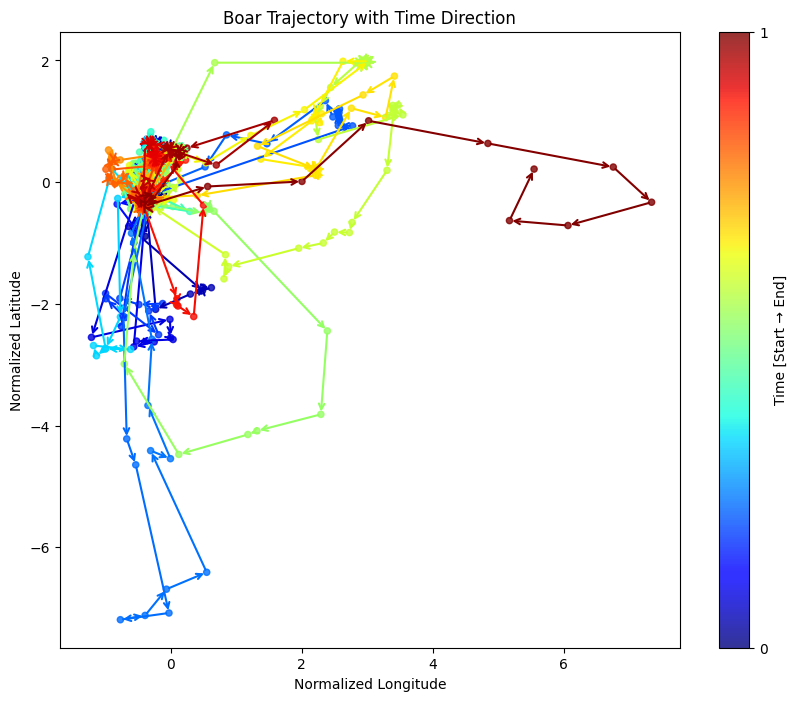}
    \subcaption{Individual trajectory over time (ID=0).}
    \label{fig:trajectory}
  \end{minipage}
  \hfill
  \begin{minipage}[c]{0.45\textwidth}  
    \centering
    \includegraphics[width=\linewidth]{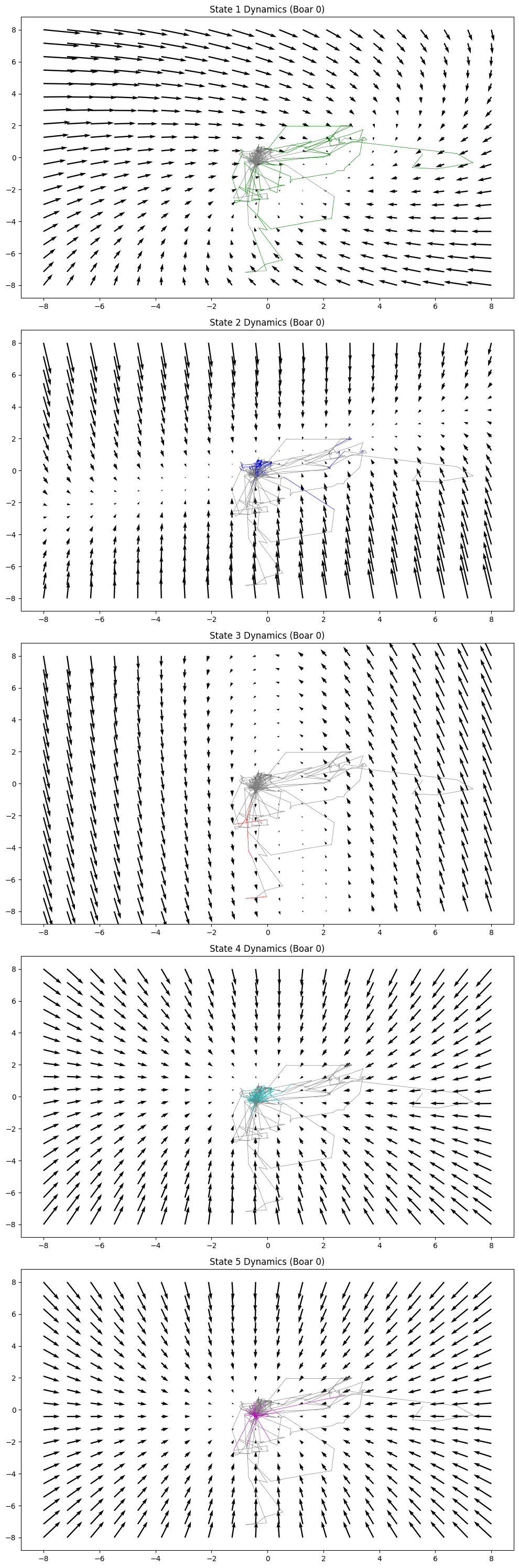}
    \subcaption{Vector fields of hidden states for the same boar.}
    \label{fig:states}
  \end{minipage}
  \caption{Example of an individual boar's movement trajectory and corresponding hidden-state vector fields.}
  \label{fig:trajectory-and-states}
\end{figure}

To improve computational efficiency while preserving essential movement patterns, we downsample the raw trajectory data (Figure~\ref{fig:trajectory}) by keeping every fifth observation (downsampling factor = $5$). Additionally, we discard the initial burn-in portion of the Markov chain to ensure convergence, retaining only samples from the stationary distribution for inference.

\subsubsection{Sensitivity analysis of AR-HMM to random seed}

To evaluate the reproducibility of inferred state sequences under different random‐seed initializations, we compute the Adjusted Rand Index (ARI) \citep{rand1971, hubert.arabie1985, warrens.vanderhoef2022} between pairs of hidden‐state trajectories. The ARI, computed as in \eqref{eq:ari}, measures the chance‐corrected agreement of all unordered pairs of time‐points: it assigns positive weight to pairs placed in the same state by both runs and negative weight when one run groups a pair together while the other does not. Formally,

\begin{equation}
\label{eq:ari}
ARI = \frac{RI - \mathbb{E}[\,RI\,]}{\max(RI) - \mathbb{E}[\,RI\,]}
\end{equation}

\noindent
where \(\mathrm{RI}\) is the classical Rand Index over the set of \(\binom{T}{2}\) time‐point pairs and \(\mathbb{E}[\mathrm{RI}]\) its expectation under random labeling. By definition, \(\mathrm{ARI}=1\) indicates perfect concordance between two runs (up to permutation of state labels), values near 0 reflect no better agreement than chance, and negative values signal systematic disagreement.

We performed 20 independent AR-HMM fits with different random seeds (see Appendix Figures~\ref{fig:appendix-ari-vs-pct-label-match} and~\ref{fig:appendix-ari-boxplot}) to assess the stability of our latent‐state inference. Pairwise comparisons using the Adjusted Rand Index yielded median ARI values above 0.75 for most individuals, demonstrating that the model consistently converges on the same state sequence regardless of initialization. Although juvenile boars showed marginally greater variability in ARI, likely a consequence of smaller sample sizes and greater early life behavioral heterogeneity, overall reproducibility remained high between sexes and all age classes. These results confirm that our autoregressive HMM framework robustly recovers latent behavioral states under different random seeds.

\subsection{Constructing the behavioral network}

Prior to constructing our targeted behavioral network, we first derive three complementary representations of wild-boar mobility. We begin with a bipartite graph $G = (B, C, E)$ in which one partition corresponds to individual boars \(B = \{b_i\}_{i=1}^N\) and the other to hexagonal grid cells \(C = \{c_j\}_{j=1}^M\) where stops were detected. An unweighted edge \((b_i,c_j)\) indicates that boar \(b_i\) recorded at least one GPS fix within cell \(c_j\).

We then project the bipartite graph into an undirected place‐level network in which two cells are connected by an edge whenever they share at least one visiting boar. Edges are unweighted, nodes two attributes derived from the full time‐series: (i) the total \emph{boar minutes} spent in that cell—computed by summing the fixed‐interval dwell times (5min per ping) of all individuals, and (ii) the total number of entry events, where for each boar we collapse any maximal, time‐contiguous block of fixes within a cell into a single visit and count a new event only when the boar departs and later re‐enters. These node‐level metrics capture both the intensity and frequency of use, setting the stage for subsequent analyses of movement structure and behavioral drivers.

Although the place‐level projection captures which nodes share visitors, it conflates all co‐occurrences into undirected links—therefore obscuring the true sequence of movements. In practice, for any boar \(b_i\) whose sequence of visits is \(c_p \to c_q \to c_r\), the projection yields a fully connected triangle among \(c_p\), \(c_q\), and \(c_r\), even though no individual ever traveled directly between \(c_p\) and \(c_r\). In other words, there should not exist a pairwise relationship between nodes \(c_p\) and \(c_r\). To resolve these misclassified relationships, we next construct a directed trajectory network in which nodes remain our hexagonal stop locations, but where edges now represent observed transitions. To be more specific, for each boar \(b_i\) we sort its GPS fixes by timestamp, assign each fix to its corresponding cell \(c_j\), collapse any consecutive fixes within the same cell into a sequence of unique node visits, and then record every observed transition from one visited cell to the next as a directed edge. We weight each directed edge by the total number of such transitions across all boars. This trajectory network therefore preserves the order and direction of movement, allowing us to pinpoint preferred routes, barriers, and the influence of environmental attractors or repellents that are invisible in an undirected co‐visitation graph.

We constructed distinct behavioral networks in the form of directed graphs for each wild boar and each state inferred by the AR-HMM. Assuming that most of the animals’ activities take place within the hexagonal cells corresponding to the nodes of the place-projected network, we used these nodes as the structural basis for building the behavioral networks.
For each state, we selected only the nodes whose corresponding hexagonal cells contained GPS points assigned to that specific state. Starting from this subset of nodes, we built a directed behavioral network by applying the same trajectory-based method used to construct the directed edges and assign the weights in the overall directed network. In this way, the weight of each edge represents the number of times the selected wild boar moved from one node to another.

\subsection{Raster data}
To understand how different natural features influence boar behavior, we assemble a 10 m-resolution covariate stack in Google Earth Engine by harmonizing five public earth observation datasets to a common grid. Categorical land cover came from the 100 meter COPERNICUS Global Land Cover 2019 \cite{Buchhorn_2020_CGLS_LC100m}. Topography was derived from the 30 m COPERNICUS digital elevation model \cite{ESA_2022_CopDEM_GLO30}, from which elevation, slope, aspect, and a 3 × 3 Terrain Ruggedness Index were calculated. The hydrological context used the Global Surface Water occurrence, recurrence, and maximum extent bands, plus a “permanent water” mask defined as > 75\% occurrence \cite{Pekel_2016_GSW_v14}. Vegetation phenology was captured by monthly median NDVI layers from Sentinel-2 \cite{ESA_2017_Sentinel2_SR}.

\subsection{Spatial modeling of mobility–entropy links}

To evaluate how local geography shapes the regularity of wild-boar movement, we treated the Shannon entropy of boar mobility within each 200-m hexagon, \(H_j\), as a quantitative response where low \(H_j\) indicates narrow, consistent activity, and high \(H_j\) marks a diversified behavioral pattern.  Every hexagonal polygon was overlaid on the 10 m raster stack, and the mean of each continuous band—elevation, slope, terrain-ruggedness index, long-term mean temperature, precipitation, and aspect—was extracted.  Aspect was re-expressed by its circular components \(\sin(\mathrm{aspect})\) and \(\cos(\mathrm{aspect})\); skewed variables were log-transformed, then all continuous covariates were centered and scaled.  

Entropy was linked to the landscape with a Gaussian spatial GAM,
\[
H_j=\alpha+\mathbf{x}_j^{\!\top}\boldsymbol{\beta}
     +f(x_j^{\mathrm{E}},x_j^{\mathrm{N}})+\varepsilon_j,\qquad
\varepsilon_j\sim\mathcal N(0,\sigma^2),
\]
where \(\mathbf{x}_j\) comprises the urban proportion together with the scaled terrain–climate predictors, and \(f(\cdot,\cdot)\) is a thin-plate spline over planar coordinates that captures residual spatial dependence.  Land-cover contrasts were gently ridge-penalized to stabilize estimates for sparsely represented classes, and all smoothing parameters were chosen by restricted maximum likelihood, yielding a parsimonious model without ad-hoc tuning.

The coefficients \(\boldsymbol{\beta}\) quantify how each geographic attribute shifts mobility entropy, whereas the spatial smooth \(f\) maps local hot- and cold-spots unexplained by the measured covariates.  This framework yields interpretable effect sizes while honoring fine-scale spatial dependence in the data.

\section{Results and discussion}

\subsection{AR-HMM model results}
\label{sec:arhmm-results}

\begin{figure}[H]
  \centering
  \includegraphics[width=0.8\linewidth]{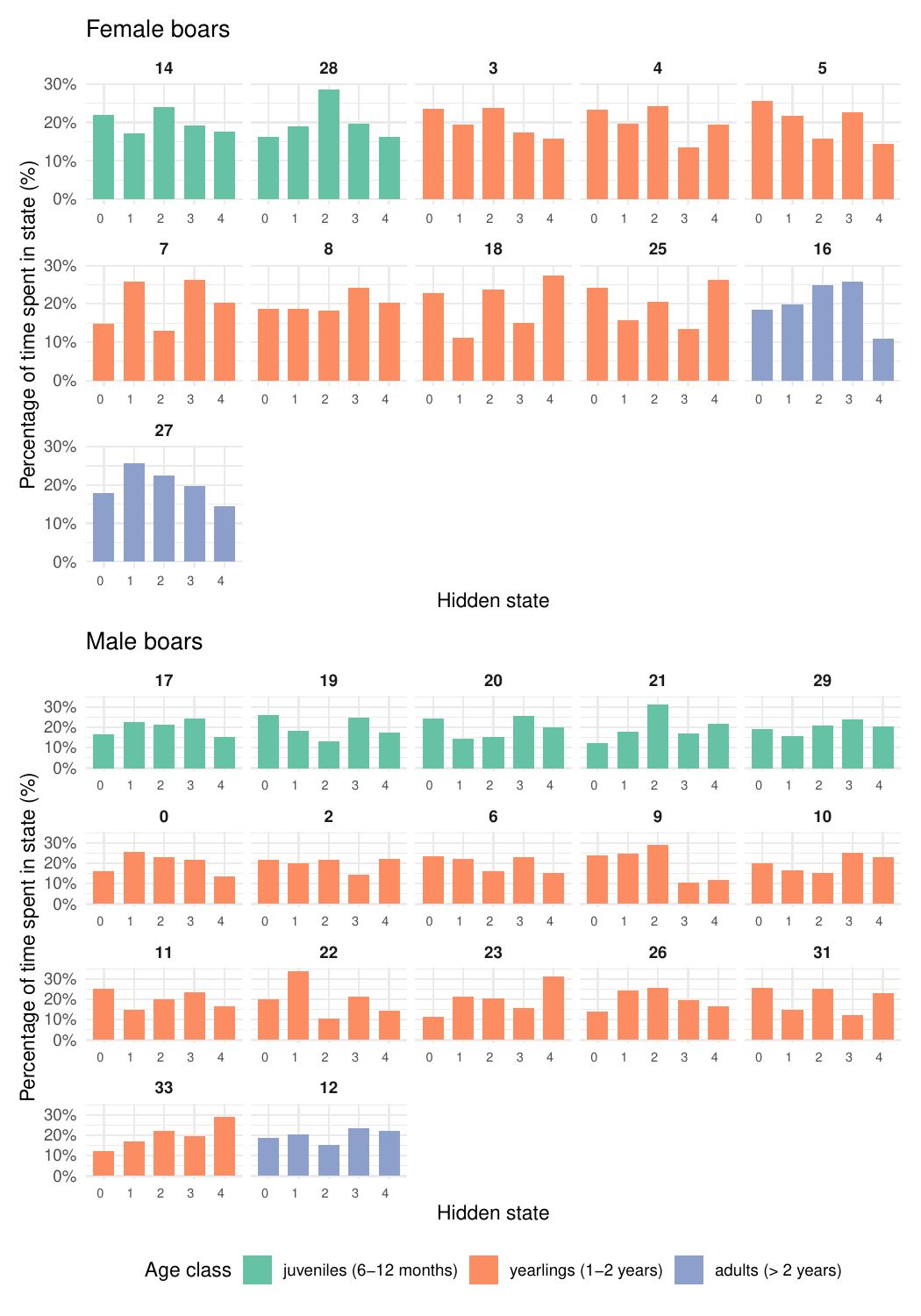}
  \caption{Results of AR-HMM Model: hidden states identified for individual boars. Panels labeled with individual boar IDs.}
  \label{fig:aslds-states-barplot}
\end{figure}

The application of the AR-HMM model to individual wild boar movement trajectories revealed distinct hidden states that characterize the animals' behavioral patterns. Figure~\ref{fig:aslds-states-barplot} presents the distribution of time spent in each hidden state, separately for female and male boars, with age classes highlighted for each individual. It is important to note that the state labels are defined independently for each boar; for example, state label 1 for boar 0 does not correspond to state label 1 for boar 18 or any other individual.

The prevalence of hidden states varied considerably between individuals, indicating substantial individual variability in movement behavior. Most individuals distributed their time relatively evenly across states, with few showing a clearly dominant state. The distribution of states appears to be more uniform in adults, while it is more varied in younger individuals. Notably, the data collection period overlaps with the breeding season, which may have influenced the diversity of movement states observed in yearlings.

\begin{figure}[H]
  \centering
  \includegraphics[width=0.8\linewidth]{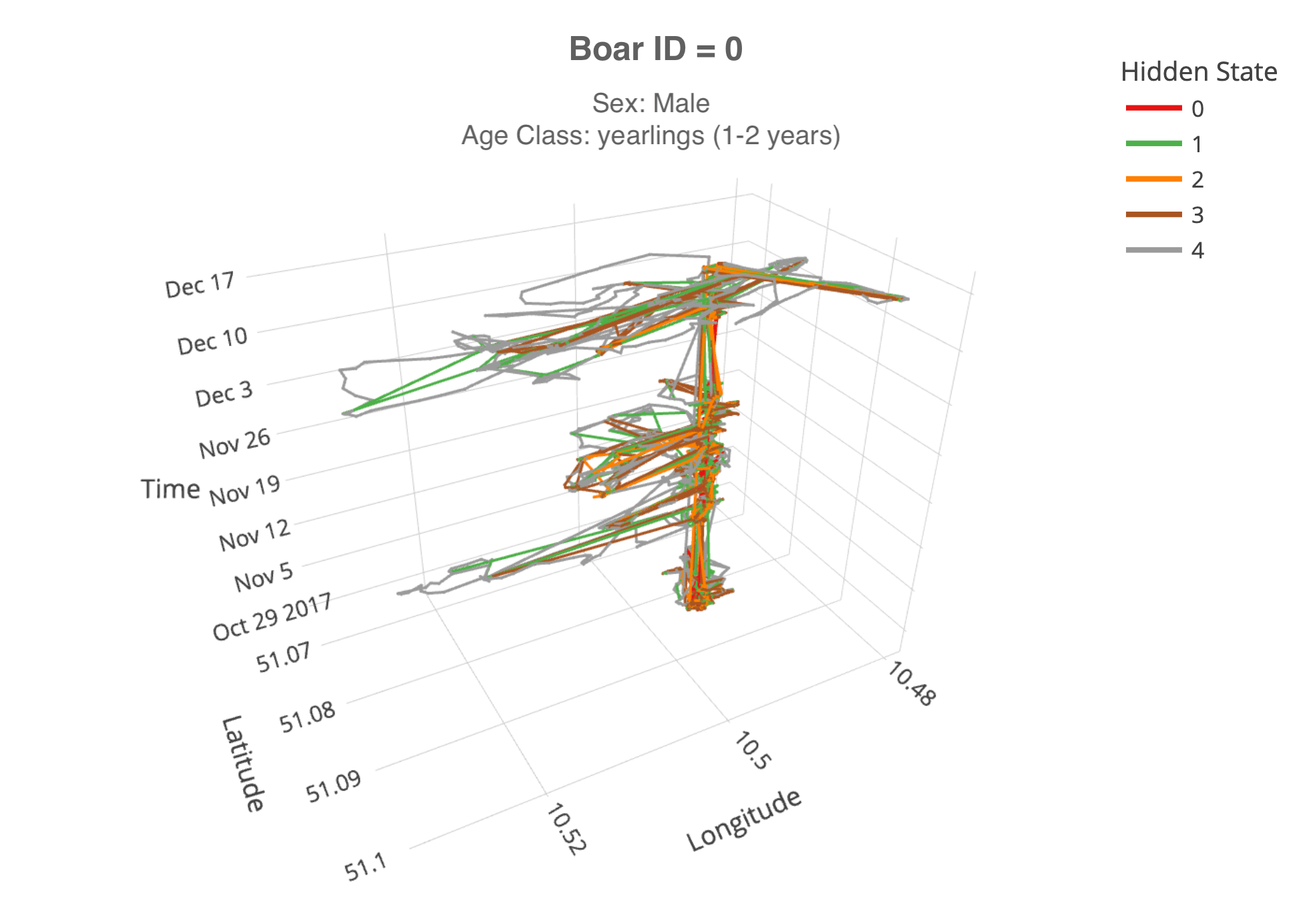}
  \caption{An example of an individual space-time trajectory labeled by hidden states.}
  \label{fig:aslds-space-time-cube-for-one-boar}
\end{figure}

In Figure~\ref{fig:aslds-space-time-cube-for-one-boar} we provide an example of an individual space-time trajectory labeled by hidden states.

Similar to our approach, a study has leveraged a semi-supervised HMM model for behavioral classification, notably in avian species (birds) \citep{saldanhaAnimalBehaviourMove2023}. This study, however, incorporated auxiliary data such as accelerometer and wet-dry sensor readings to inform their semi-supervised learning and assign specific behavioral labels to the identified states. In contrast, our study lacked such ground truth data for direct state classification. Consequently, while our HMM successfully identified 5 distinct behavioral states within the boars movement data, the precise interpretation or labeling of these states (e.g., foraging, resting, traveling) remains an area of future investigation as we lack ground truth data, and we cannot assign labels to our identified states.

\subsection{Descriptive space and trajectory statistics}
\begin{figure}[!ht]
  \centering
  \begin{subfigure}[b]{0.45\linewidth}
    \includegraphics[width=\linewidth]{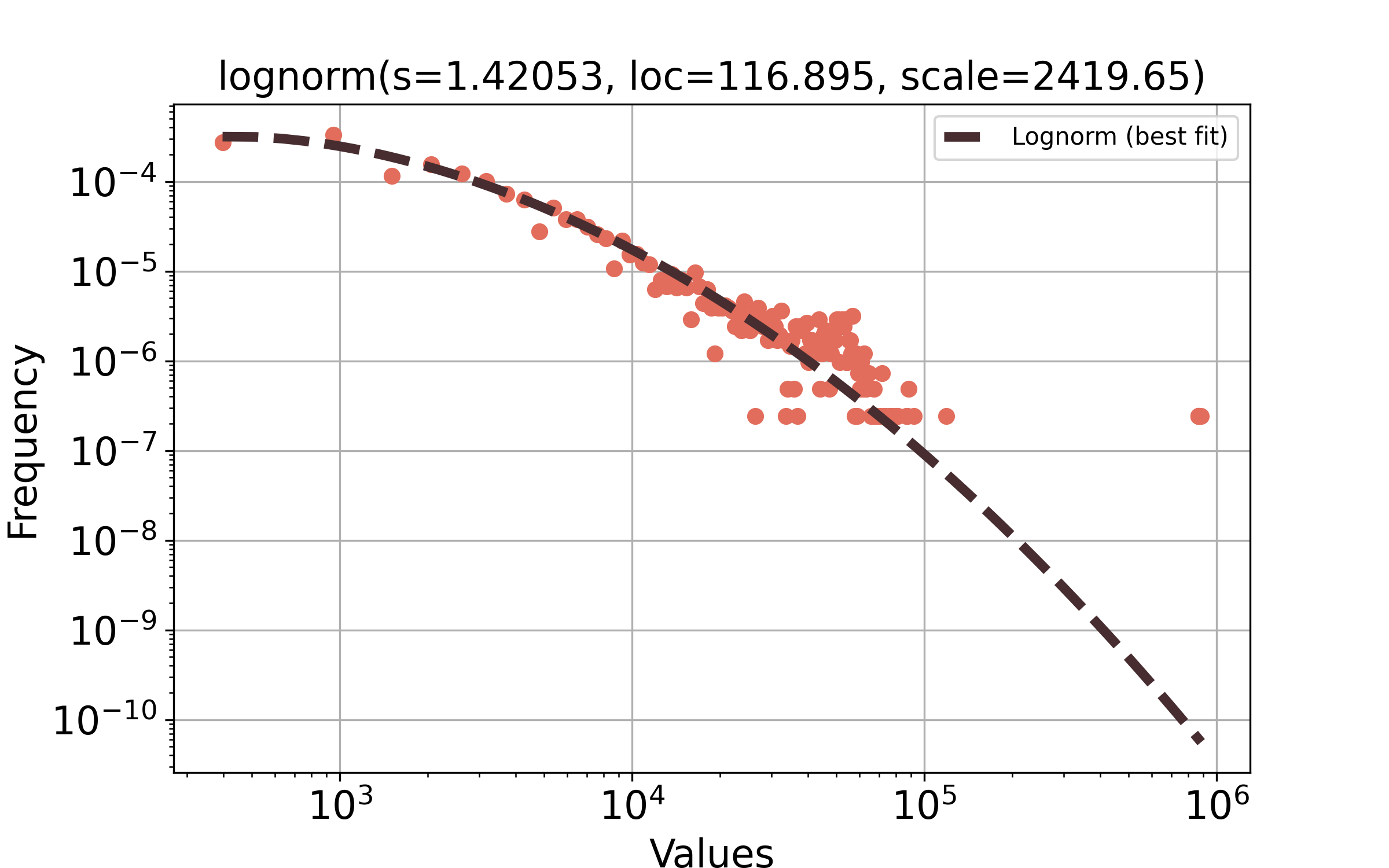}
    \caption{Waiting times.}
    \label{fig:waiting_times}
  \end{subfigure}
  \hfill
  \begin{subfigure}[b]{0.45\linewidth}
    \includegraphics[width=\linewidth]{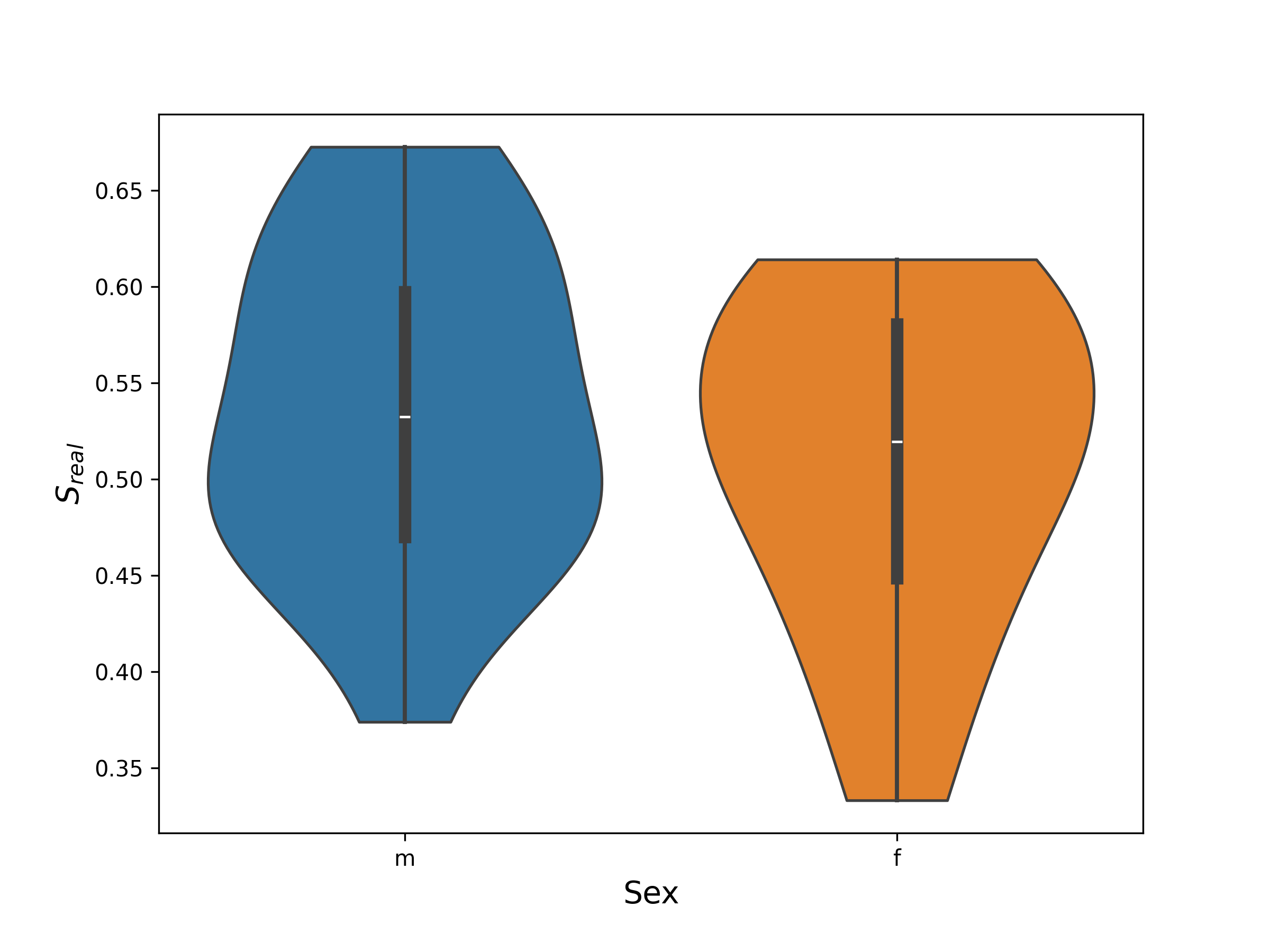}
    \caption{Predictability by sex.}
    \label{fig:predictability}
  \end{subfigure}

  \vspace{1em}

  \begin{subfigure}[b]{0.45\linewidth}
    \includegraphics[width=\linewidth]{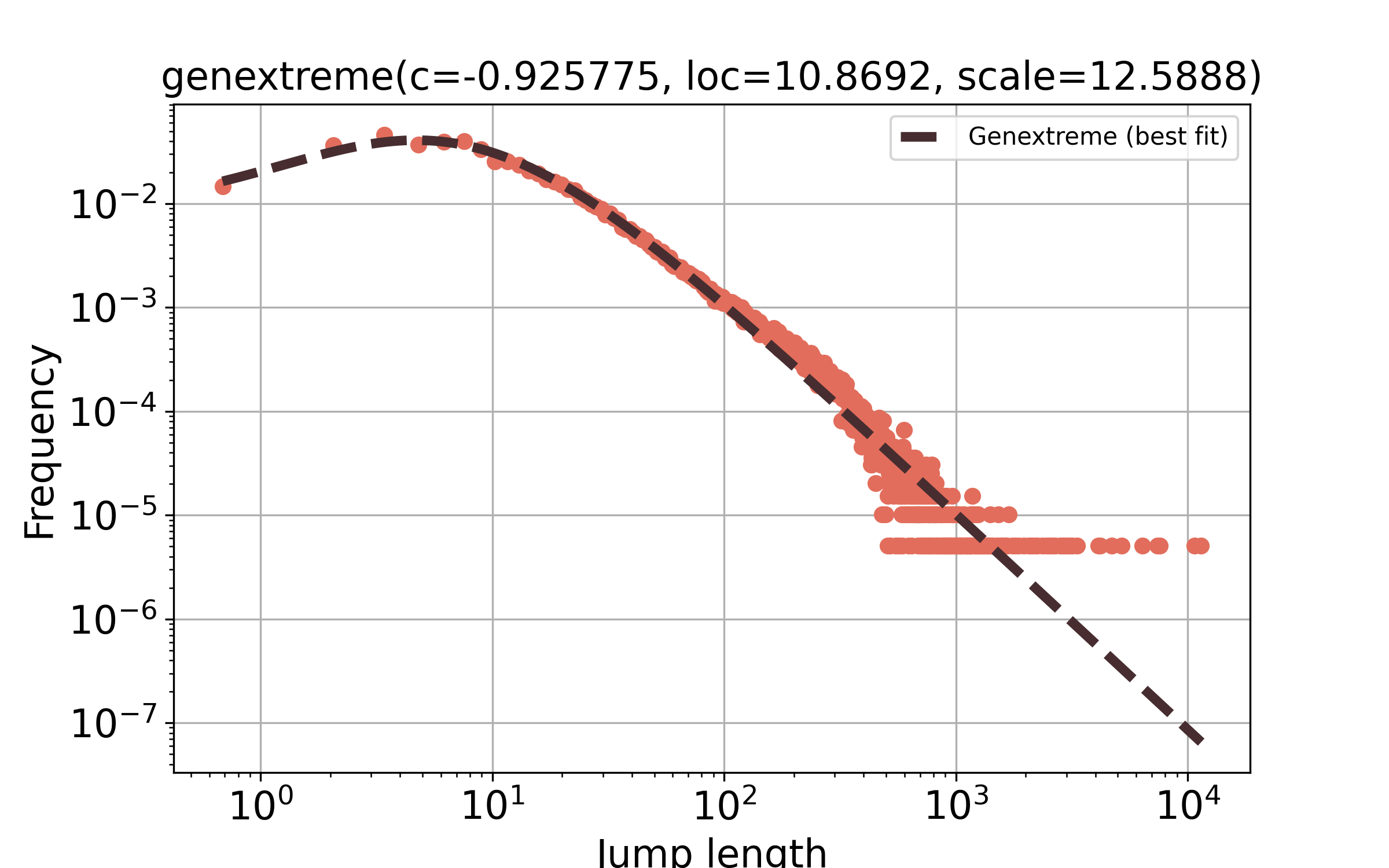}
    \caption{Jump length distribution.}
    \label{fig:jump_distribution}
  \end{subfigure}
  \hfill
  \begin{subfigure}[b]{0.45\linewidth}
    \includegraphics[width=\linewidth]{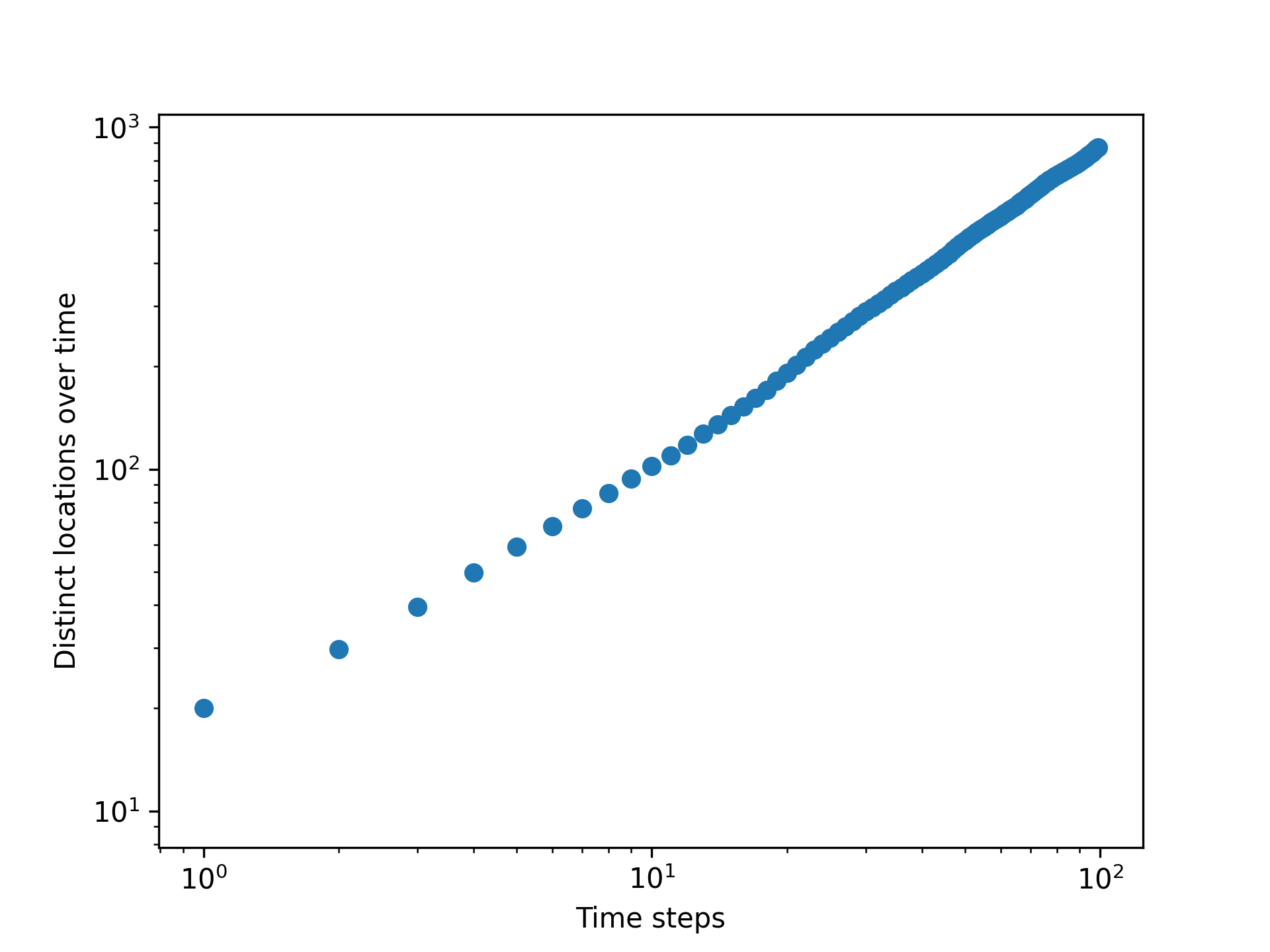}
    \caption{Distinct locations over time.}
    \label{fig:distinct_locations}
  \end{subfigure}

  \vspace{1em}

  \begin{subfigure}[b]{0.45\linewidth}
    \includegraphics[width=\linewidth]{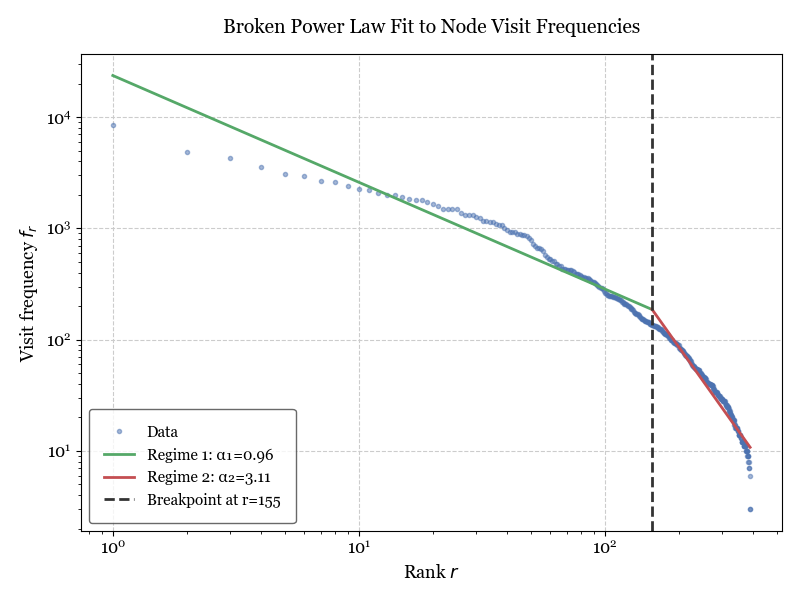}
    \caption{Node visit frequency.}
    \label{fig:visit_frequency}
  \end{subfigure}

  \caption{\small Descriptive statistics summarizing key properties of wild boar movement.}
  \label{fig:five-graphs}
\end{figure}

Movement of wild boars, analyzed through the lens of mobility statistics (Figure~\ref{fig:five-graphs}), reveals regularities and characteristic patterns in their movement behavior. The distribution of waiting times (Figure~\ref{fig:waiting_times}) is well-approximated by a log-normal distribution, indicating that short stops are most probable, while few stops dominate the overall waiting time. Movement predictability differs between sexes (Figure~\ref{fig:predictability}), with males being slightly more predictable than females. The distribution of jump lengths (Figure~\ref{fig:jump_distribution}) highlights occasional long-distance displacements, though short trips are most frequent in wild boar movement. The number of distinct locations visited over time (Figure~\ref{fig:distinct_locations}) grows sublinearly, suggesting a tendency toward preferential return and limited spatial exploration. This is further emphasized by the node visit frequency (Figure~\ref{fig:visit_frequency}), which follows a broken power-law distribution, indicating heterogeneous space use, with a few highly visited locations and many rarely visited ones.

\subsection{Network analysis}
To understand how wild boars interact and which individuals are pivotal to the spatial‐interaction network, we conduct a targeted percolation analysis—also known as \emph{attack percolation}.  Starting from the full bipartite graph \(G_{UH}\) whose two partitions are the set of boars \(U\) and stop locations \(H\), we iteratively delete boar vertices in descending order of their degree \(k_{u}\).  After each deletion step, we project the surviving bipartite graph onto a weighted, undirected stop-stop graph \(G_{H}\), where the weight \(w_{ij}\) on edge \((i,j)\) equals the summed intensity of shared visits.  We then measure the size of the \emph{giant connected component}, \(S_{\mathrm{GCC}}\), and track it as a function of the cumulative number of boars removed, \(k\).  The resulting percolation curve quantifies the robustness of landscape connectivity and reveals those boars whose removal precipitates a sharp decline in \(S_{\mathrm{GCC}}\), identifying them as critical movement ``bridges''. We repeat this process using stops instead of boars to see which areas are critical for the boar-boar social network, informing decisions about disease spread.

\begin{figure}[!ht]
    \centering
    \includegraphics[width=0.8\linewidth]{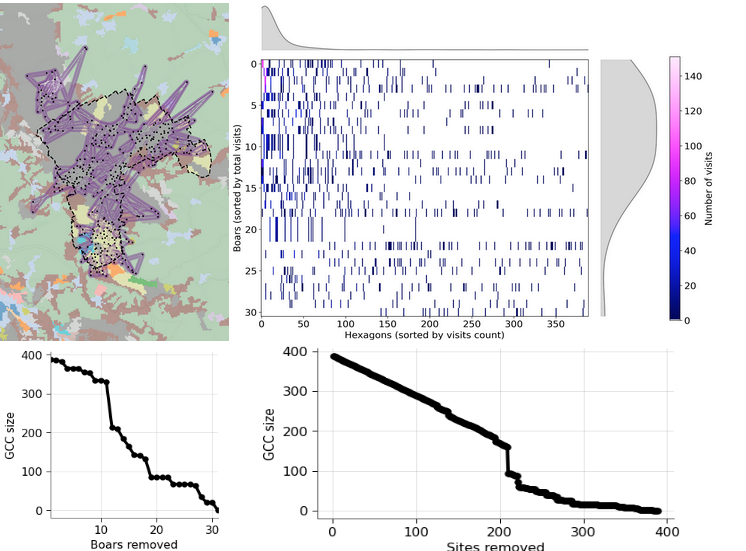}
    \caption{Network analysis. We show the path network of all boar trajectories and their stops in the top left panel, along with a heatmap of the sites visited by all boars. beneath we show boar percolation and site percolation results. There is a significant transition indicating that one node is crucial for the connectivity of the network.}
    \label{fig:network_fig}
\end{figure}

Figure \ref{fig:network_fig} shows the network built from the trajectories and stops of all boars. Edges are weighted the amount of times all boars traveled between two stops. Nodes attributes are the amount of minutes spent by all boars at the stop, and the number of visits by all boars. These attributes serve as a measure of node relevance. The right panel shows a heatmap of the visits per boar per node.

\begin{figure}[!ht]
  \centering
  \includegraphics[scale=0.4]{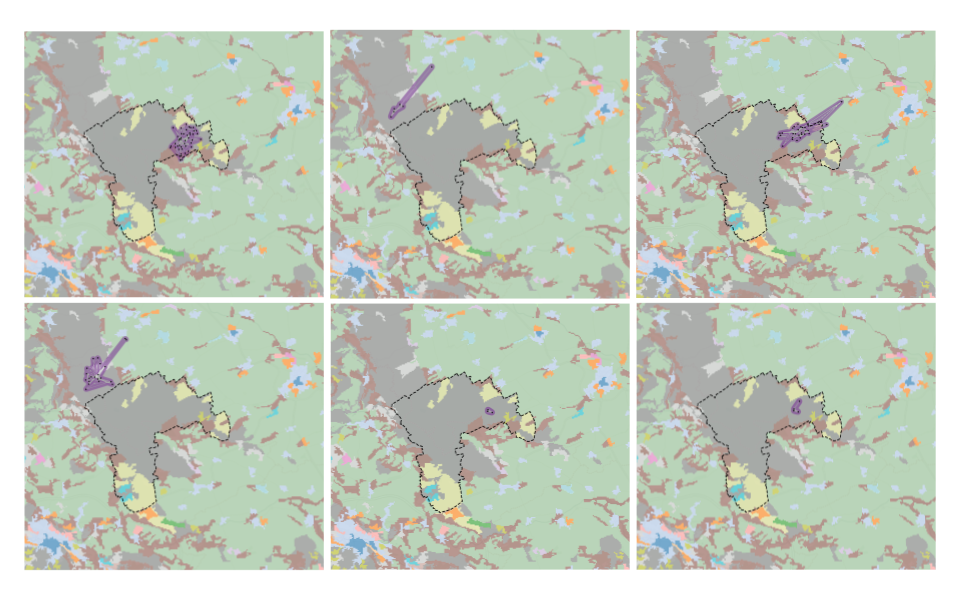}
  \caption{Behavioral networks for two wild boars across different states inferred by the AR-HMM. Each row corresponds to a different individual wild boar, and each column represents a distinct behavioral state.}
  \label{fig:behavior_networks}
\end{figure}

Figure \ref{fig:behavior_networks} shows the directed behavioral networks constructed for two wild boars across the different states inferred by the AR-HMM. Each row corresponds to a different individual, and each column to a specific state. Nodes represent hexagonal cells where GPS points were detected for the given state, and directed edges indicate transitions between nodes, with edge weights proportional to the number of times the boar moved from one cell to another.

\subsection{Spatial entropy}

Behavioral entropy proved remarkably stable across the study area: most hexagons were dominated by homogeneous mobility patterns, and overall variability in the index was modest.  Where differences did emerge, they were chiefly attributable to land cover.  Hexagons whose modal class was mixed forest showed the clearest increase in entropy (\(+0.36 \pm 0.12\), \(p = 0.003\)), consistent with a wider behavioral repertoire in structurally heterogeneous woodland.  Pasture and conifer forest displayed weaker, marginally significant rises, whereas elevation, slope, terrain ruggedness, temperature, precipitation, and aspect had no detectable effect once land cover was taken into account.  

A highly significant spatial smooth (\(F = 3.05\), \(p < 2 \times 10^{-16}\); 28 effective degrees of freedom) indicated that pockets of unusually low or high entropy persist beyond the measured covariates, clustering near the park–settlement interface and within central forest blocks.  Land cover together with this residual spatial field explained 39\,\% of the between-hexagon deviance (\(R_{\mathrm{adj}}^{2} = 0.32\)), implying that unmeasured factors—such as hunting intensity or mast dynamics—rather than broad topography or macro-climate account for the remaining variation in behavioral regularity.

\begin{figure}[H]
  \centering
  \includegraphics[width=0.7\linewidth]{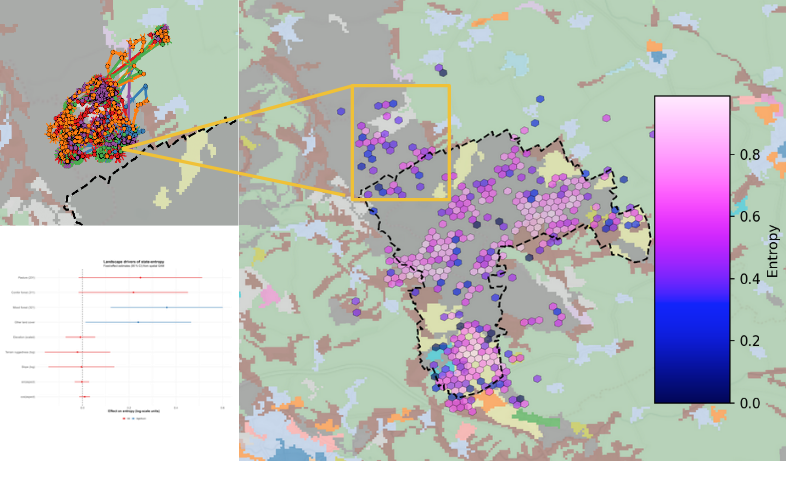}
  \hfill

  \label{fig:entropy}

  \caption{The entropy of hidden states calculated over the tesselation. the top left panel shows a focus of the behaviors of a single boar.}
\end{figure}

\subsection{Boars' movements and African Swine Fever}

Wild boars are a central key to the epidemiology and persistence of African Swine Fever (ASF), acting as both reservoirs and vectors for the highly virulent virus \citep{pollockPredictingHighriskAreas2021}. The dynamics of ASF transmission are complex, involving both direct contact between live animals and, crucially, indirect transmission through contaminated environments and infectious carcasses. While direct transmission, occurring through close contact between infected and susceptible wild boars, is a recognized pathway \citep{guinatTransmissionRoutesAfrican2016}, its overall contribution to large-scale disease spread appears to be inherently limited by the severe pathology of ASF \citep{podgorskiWildBoarMovements2018}. The high lethality of the virus, coupled with a relatively short infectious period ranging from 1 to 6 days \citep{yoonEstimatingTimeInfection2023}, and the inherent social structure of wild boar populations, often restricts direct transmission to immediate group members. This means that while individual wild boar movements are essential for the initial encounter, the disease’s rapid progression can quickly hamper extensive movements of infected individuals, thereby mitigating their role in long-distance viral dissemination \citep{podgorskiWildBoarMovements2018, oneillModellingTransmissionPersistence2020}. In contrast, indirect transmission emerges as a predominant and persistent driver of ASF outbreaks and long-term viral maintenance within wild boar populations. Infected carcasses and the environments they contaminate serve as critical sources of the virus \citep{podgorskiWildBoarMovements2018,chenaisIdentificationWildBoar2018, sauter-louisAfricanSwineFever2021, limModelingSingaporesFirst2024}. The remarkable stability of the ASF virus, allowing it to survive for extended periods -potentially months, withing carcasses, means that these remains can act as reservoirs, enabling new infections long after the initial one \citep{probstBehaviourFreeRanging2017, fischerStabilityAfricanSwine2020, gervasiAfricanSwineFever2021, arzumanyanPossibilityLongtermSurvival2021}.

This environmental persistence is particularly significant, as carcass-mediated transmission can account for a substantial proportion of new cases, effectively decoupling disease spread from the extensive movements of live infected animals. Modeling and field studies have shown that a substantial proportion of new ASF cases—often over half—can be attributed to contact with infectious carcasses, rather than direct transmission between live animals, especially in low-density populations where direct contact is less frequent \citep{langeElucidatingTransmissionParameters2017, pepinEcologicalDriversAfrican2019, gervasiAfricanSwineFever2021}. Similar mechanisms have been observed in other diseases, such as anthrax, where carcass sites create locally infectious zones that attract susceptible animals and facilitate indirect transmission \citep{turnerLethalExposureIntegrated2016, walkerUngulateUseLocally2020}.

This nuanced understanding of ASF transmission pathways highlights the profound significance of identifying and managing attractors and repellents in the landscape, particularly through the lens of movement-based behavioral networks. Attractors, in this context, are areas that consistently draw wild boars due to the presence of vital resources of favorable conditions. These can include prime foraging grounds, communal resting sites, or water sources. Such location becomes hotspots for both direct and indirect transmission. They facilitate increased contact rates among wild boars, but more critically, they are areas where infected carcasses are likely to accumulate, and where environmental contamination with the virus is concentrated. \citet{cukorWildBoarCarcasses2024} proved that an enormous wild boar carcass is highly attractive to other boars. Therefore, identifying these attractors is crucial for targeted surveillance, effective carcass removal strategies \citep{gervasiCombiningHuntingIntensive2022}, and the implementation of biosecurity measures. Conversely, repellors are areas that wild boars actively avoid, such as human settlements, busy roads, or unsuitable habitats. These features can act as natural or anthropogenic barriers to movement. While their primary role is to deter wild boar presence, their identification is equally crucial for disease management and containment. By understanding what repels wild boars, strategic interventions, such as the placement of physical barriers or the modification of landscapes such as landscape fragmentation \citep{vargas-amadoPotentialEffectManaging2023,gelmi-candussoDynamicConnectivityAssessment2024}, can be designed to limit their movement into vulnerable areas or to create effective disease containment zones \citep{salazarEffectsHabitatFragmentation2022}. Animal movement trajectories can be analyzed to identify frequently used movement pathways or corridors. These corridors often connect different resource patches (attractors) and represent areas of high movement intensity \citep{faustiniHabitatSuitabilityMapping2025}. This approach enables the development of more accurate predictive models, dynamic risk mapping, and the identification of key attractors or groups of individuals that disproportionately contribute to transmission. Future research should focus on refining these analytical tools, quantifying the strength of spatial preferences and investigating the impact of management interventions on wild boar movement behavior. Ultimately, leveraging these advanced ecological insights will be crucial for developing more effective, adaptive and spatially explicit strategies to control African Swine Fever and protect both wild and domestic pig populations.

\section{Conclusions}

In this paper, we attempt to link latent mobility behaviors constructed via a hidden Markov model with geographic characteristics and network structure. Our analysis of behavioral entropy reveals a significant insight into the spatial distribution of boars' behaviors. We observe that most nodes within the network exhibit high behavioral entropy, indicating a relatively uniform distribution of various behaviors across these locations. Conversely, only a few nodes showed low entropy, suggesting that specific behaviors are more concentrated in these limited areas. This finding implies that, for the most part, boars engage in a diverse range of activities throughout their habitat. Furthermore, individual boar behavior analysis suggests that some boars have specific expansive or spatially stable behavior. Our approach doesn't allow us to identify key distinctions in latent states both (a) between boars and (b) within boars. As a result, our state interpolations are largely uninformative on the relationship between boar behavior and geographic attributes. 

A key interpretation of these results is that the hidden behaviors identified by the model appear to be largely independent of geographical locations. If the model accurately reflects actual boar behaviors, this would suggest that boars display a wide array of behaviors in most locations, with only a few locations being associated with a singular specific behavior. This outcome is considered plausible, particularly given that a substantial portion of the Hainich National Park consists of a uniform forest environment where the boars can roam undisturbed. Furthermore, the study noted that entropy tends to be higher within the forest areas and decreases as one approaches the edges of the park. 
Finally, our network analysis is an important starting point in suggesting surveillance procedures to limit the spread of disease among wild boars and from wild boars to domestic farms. We performed a percolation analysis that shows that specific boars can be crucial for the connectivity of the network. This suggests that monitoring important boars could do more to limit the spread of disease than monitoring locations, which turn out to be only moderately significant.

\section{Limitations}

Reproducibility of the latent–state assignments was generally high (median ARI >0.75; Figs.~\ref{fig:appendix-ari-vs-pct-label-match}–\ref{fig:appendix-ari-boxplot}), yet notable differences emerged across demographic strata.  The yearling cohorts—both female and male—displayed the broadest dispersion of ARI values, with several individuals falling below 0.40. Because yearlings constitute the largest subsamples in our data set, this spread is unlikely to result from sampling artifacts and instead suggests genuine heterogeneity in yearling movement behavior. Juveniles and adults, by contrast, are represented by fewer individuals; in these smaller cohorts a single atypical trajectory can exert disproportionate influence on the group‐level stability metric, so their observed variability may partially reflect sparse sampling.  At the same time, it is improbable that the limited juvenile and adult samples fortuitously captured only the most predictable boars.  Moreover, existing ethological evidence indicates that yearlings typically exhibit higher activity levels, adults show reduced movement, and juveniles often affiliate closely with older kin.  The stability of our model, maintained across these ontogenetic activity gradients, provides additional support for the robustness of the state-detection framework.

The stability analysis employed 20 random initializations.  Although this revealed no systematic drift in label assignments, it nonetheless explores only a fraction of the model’s initial‐condition space.  Alternative starting points (particularly those that place greater weight on rare movement motifs) might converge on substantively different state partitions.  Future work should therefore extend convergence diagnostics to a broader range of initial conditions and, where feasible, adopt fully Bayesian posterior sampling to propagate this residual uncertainty.

All experiments fixed the number of latent classes at K=5.  This heuristic aligns with earlier movement‐ecology studies and yields an interpretable behavioral partition, yet it imposes an a priori level of granularity.  Rarer or transient states may have been subsumed into dominant classes, potentially obscuring ecologically meaningful nuance.  Sensitivity analyses across alternative K values, or preferably non-parametric Bayesian extensions that enable the data to determine the required number of states, would clarify whether finer behavioral distinctions modify the principal conclusions.

\section{Acknowledgements}
This work is the output of the Complexity72h workshop, held at the Universidad Carlos III de Madrid in Leganés, Spain, 23-27 June 2025. https://www.complexity72h.com

\bibliographystyle{unsrtnat}
\bibliography{references.bib}  






\clearpage
\appendix

\section{Appendix}
\label{sec:appendix-figures}

\begin{figure}[ht]
  \centering
  \includegraphics[width=1\linewidth]{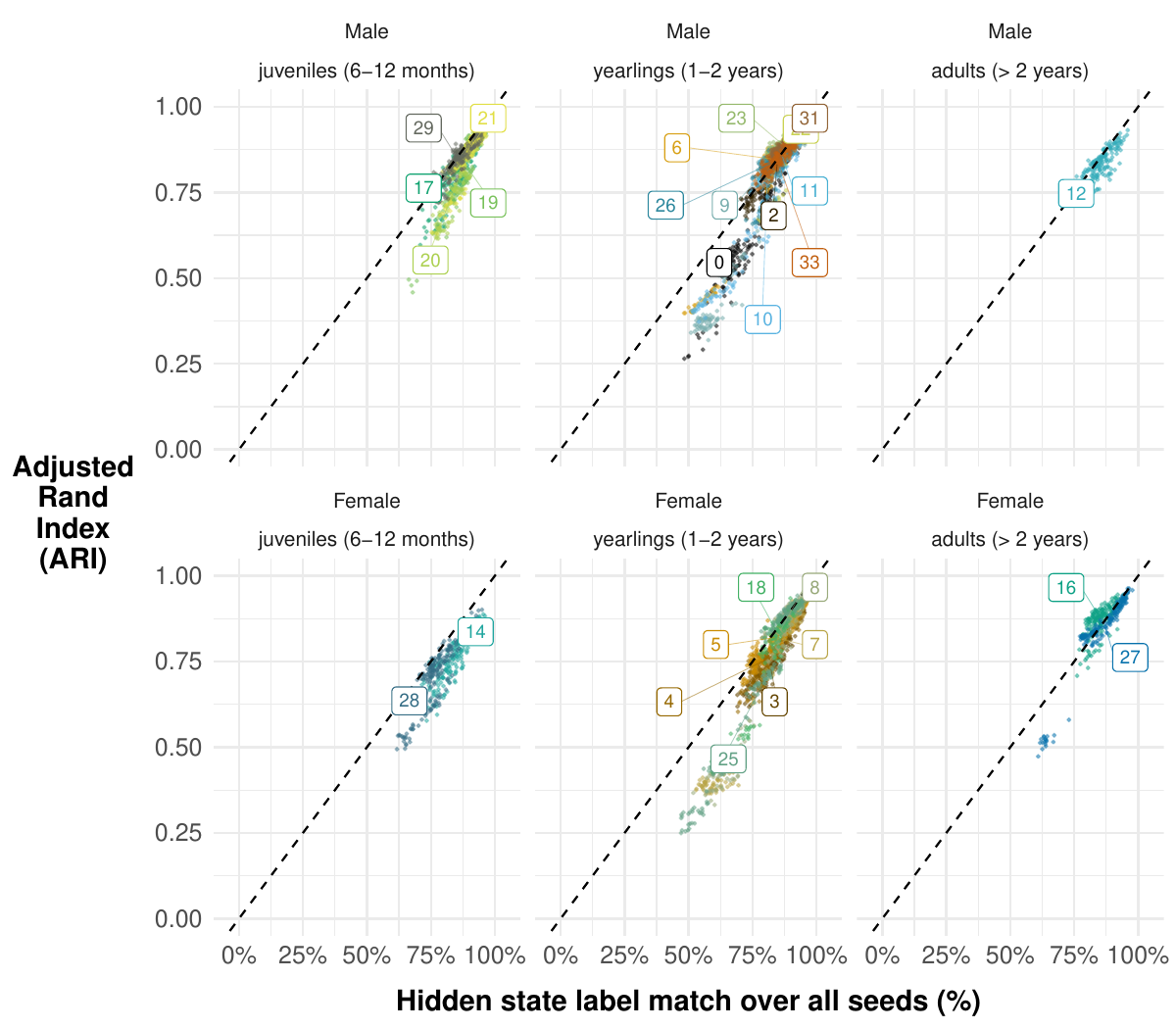}
  \caption{Stability of latent behavioral‐state detection across random seeds. Scatter plots of Adjusted Rand Index (ARI) versus the percentage of hidden‐state label agreement (over 20 runs with different random seeds) for male (top row) and female (bottom row) wild boars, separated into juveniles (6–12 months), yearlings (1–2 years), and adults (> 2 years). Each colored point represents one latent state (annotated by its state ID), and the dashed diagonal line marks perfect correspondence (ARI = label‐match). High clustering agreement (most points lying near the top‐right quadrants) indicates that our autoregressive HMM reliably identifies the same behavioral states regardless of initialization, with slightly greater variability observed in yearlings subgroups. Sample sizes in some age–sex cohorts remain modest, which could in principle exaggerate apparent stability. However, the consistent reproducibility observed in both the youngest and oldest cohorts substantially reduces the likelihood that these results are mere artefacts of random initialization. Moreover, existing ethological evidence suggests that yearlings typically exhibit higher activity levels, whereas adults display reduced movement, and juveniles often affiliate closely with their elder kin. The fact that our model’s stability persists across these ontogenetic activity gradients further reinforces the robustness of our state-detection framework.}
  \label{fig:appendix-ari-vs-pct-label-match}
\end{figure}

\begin{figure}[ht]
  \centering
  \includegraphics[width=1\linewidth]{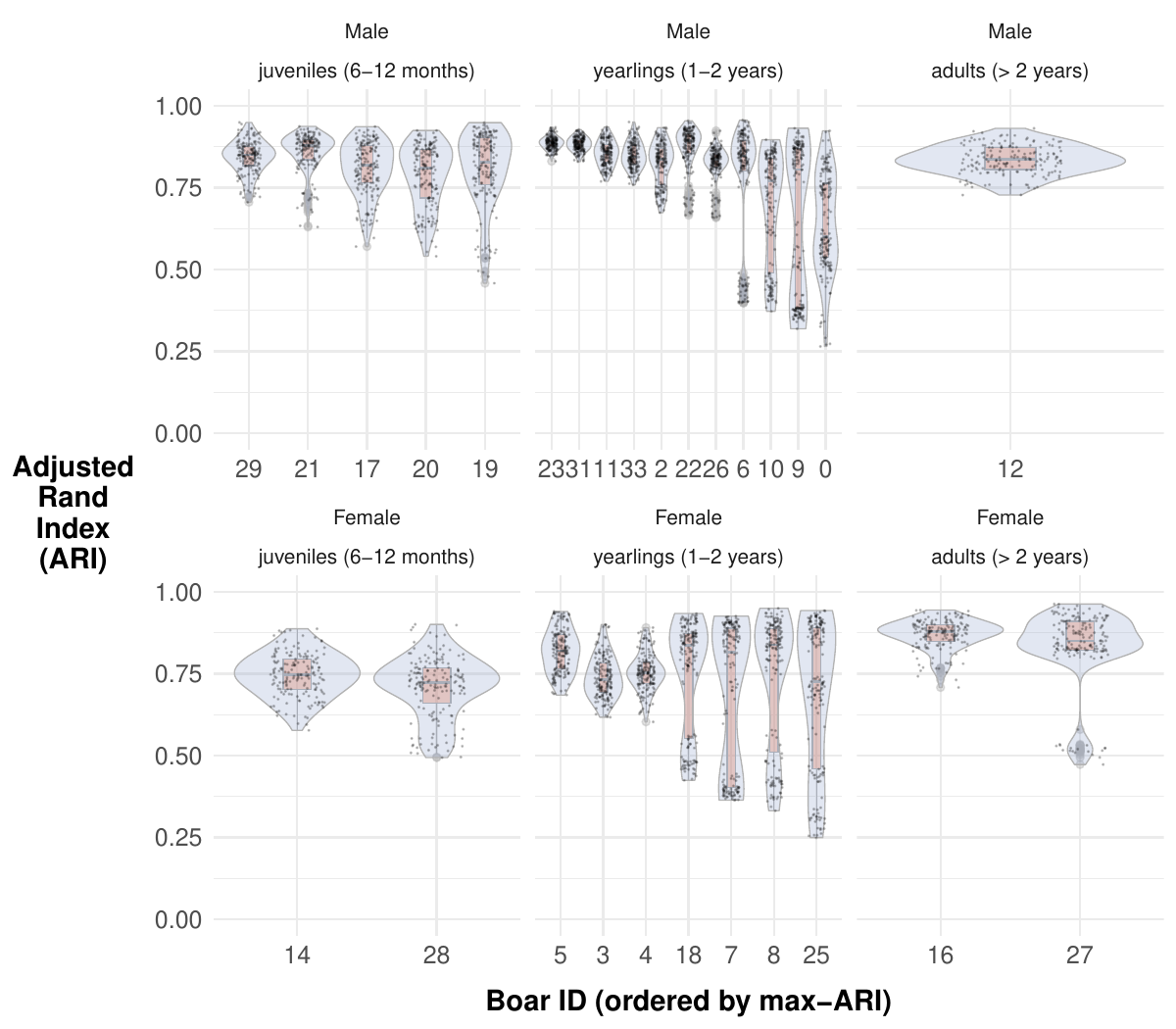}
  \caption{Consistency of latent-state assignments across model runs, by individual boars. For each boar (x-axis, ordered by highest to lowest reproducibility of labels), the shaded contours show the spread of agreement scores (Adjusted Rand Index) over 20 independent model fits, while the central box highlights the median and interquartile range of percentage label-match. Panels are arranged by sex (rows) and age class (columns). Narrow, high-centered contours indicate individuals with very stable state assignments regardless of initialization; wider or lower-lying contours reveal greater sensitivity to random seed. Across all demographic groups, most boars achieve median agreement above 0.75, demonstrating the reliability of our AR-HMM in capturing consistent behavioral patterns.}
  \label{fig:appendix-ari-boxplot}
\end{figure}

\end{document}